\documentclass{emulateapj}

\slugcomment{Accepted by ApJ}
\shorttitle{NIR Imaging of the Antennae Galaxies}
\shortauthors{Brandl et al.}

\begin{document}

\title{Deep Near-Infrared Imaging and Photometry of the Antennae
       Galaxies with {\sl WIRC}}


\author{B.R. Brandl}
\affil{Leiden Observatory, P.O. Box 9513, 2300 RA Leiden, The Netherlands}
\email{brandl@strw.leidenuniv.nl}
\author{D.M. Clark, S.S. Eikenberry} 
\affil{Department of Astronomy, University of Florida, Gainesville, 
       FL 32611}
\author{J.C. Wilson} 
\affil{University of Virginia, Department of
       Astronomy, P.O. Box 3818, Charlottesville, VA 22903}
\author{C.P. Henderson, D.J. Barry, J.R. Houck}
\affil{Cornell University, Center for Radiophysics and Space Research, 
       Ithaca, NY 14853}
\author{J.C. Carson} 
\affil{JPL, Earth \& Space Science, Pasadena, CA 91109 \& Caltech, 
       Pasadena, CA 91125}
\and
\author{T.L. Hayward}
\affil{Gemini Observatory, Southern Operations Center, Casilla 603
       La Serena, Chile }

\begin{abstract}
We present deep near-infrared images of the Antennae galaxies, taken
with the Palomar Wide-Field Infrared Camera {\sl WIRC} The images
cover a $4.\!'33\times 4.\!'33\ (24.7\mbox{\,kpc}\times
24.7\mbox{\,kpc}$) area around the galaxy interaction zone.  We derive
$J$ and $K_s$ band photometric fluxes for 172 infrared star clusters,
and discuss details of the two galactic nuclei and the overlap region.
We also discuss the properties of a subset of 27 sources which have
been detected with WIRC, {\sl HST} and the VLA.  The sources in common
are young clusters of less than 10 Myr, which show no correlation
between their infrared colors and 6~cm radio properties.  These
clusters cover a wide range in infrared color due to extinction and
evolution.  The average extinction is about $A_V\sim 2$~mag while the
reddest clusters may be reddened by up to 10 magnitudes.
\end{abstract}



\section{Introduction}

The Antennae galaxies, NGC\,4038/39 (Arp\,244), are probably the
best-known example of a pair of interacting galaxies.  At a distance
of only 19.2~Mpc\footnote{assuming $H_0 = 75\mbox{km}^{-1}
\mbox{s}^{-1} \mbox{Mpc}^{-1}$} \citep{whi99} the Antennae system has
been thoroughly studied over a large range of wavelengths.
 
Numerous observations that cannot be listed here individually have
been made at far-infrared, sub-millimeter and radio wavelengths.  They
generally agree that the most of the emission at longer wavelengths
comes from the highly extincted overlap region.  The largest molecular
complexes have masses of $(3-6)\times 10^8 M_\odot$, typically an
order of magnitude larger than the largest structures in the disks of
more quiescent spiral galaxies \citep{wil00}.  These authors also
found an excellent correlation between the CO emission and the
$15\mu$m emission seen by {\sl ISO} \citep{mir98}.  Recent mid-IR
observations at slightly higher spatial resolutions with {\sl Spitzer}
\citep{wan04} showed that the rate of star formation per unit mass in
the active areas is comparable to those in starburst and some
ultra-luminous galaxies.

The first deep optical analysis of the Antennae with the Wide Field
Camera on {\sl HST} \citep{whi95} showed over 700 point-like objects.
Subsequent observations with {\sl WFPC2} \citep{whi99} increased the
sensitivity by 3 magnitudes in {\sl V} band and revealed between 800
and 8000 clusters in four age ranges: {\sl (i)} ages of $\leq 5$~Myr
around the edges of the overlap region and $5-10$~Myr in the western
loop, {\sl (ii)} ages $\sim 100$~Myr in the northeastern star
formation region, {\sl (iii)} intermediate-age clusters of $\sim
500$~Myr and {\sl (iv)} old globular clusters from the progenitor
galaxies.  While \citet{whi99} and \citet{fri99} studied the
statistical properties of the cluster population, \citet{gil00} and
\citet{men02} investigated the properties of selected ``super star
clusters'' in greater detail.

X-ray observations with {\sl Chandra} \citep{zez02} revealed 49
sources, including several ultra-luminous X-ray (ULX) sources
with X-ray luminosities of $L_X > 10^{39}$\,erg\,s$^{-1}$, suggesting
these are binary accretion sources.

So far, most studies of the star clusters in the Antennae have been
focused on a single wavelength regime with few exceptions:
\citet{zha01} studied the relationship between young star clusters and
the interstellar medium based on observations ranging from X-rays to
the radio wavelengths, and \citet{whi02} correlated optically detected
star clusters with their radio counterparts from
\citep{nef00}. \citet{kas03} combined UBVRJHK images to derive
extinction maps for the Antennae and found several red clusters.

This paper is the first one of a series to study the correlation
between the near-infrared sources based on the {\sl WIRC} images and
their radio counterparts (\citet{nef00}, \citet{whi02}); subsequent
papers \citep{cla05} focus on the correlation between the infrared and
X-ray sources \citep{zez02}.  Here we present deep near-infrared
observations of NGC\,4038/39 obtained with {\sl WIRC}.  We give a
brief description of the instrument, discuss the infrared morphology
of the Antennae galaxies, and our photometric methods.  In
section~\ref{secdiscussion} we discuss the properties of the infrared
sources and focus on the infrared sources with radio and optical
counterparts.


\section{The Data}
\subsection{Observations}
\label{secobservations}

We observed NGC\,4038/39 with the new wide-field infrared camera {\sl
WIRC} \citep{wil03} mounted to the prime focus of the $200''$ Hale
telescope on Palomar Moutain.  The observations were part of the
commissioning of the instrument.  The ``original''
version\footnote{Recently, a collaborative agreement with Caltech
allowed the upgrade to a $2048\times 2048$ pixel$^2$ Rockwell
Hawaii-II array that provides a $8.\!'7\times 8.\!'7$ field of view.}
of {\sl WIRC}, which was used for the observations presented in this
paper, featured a field-of-view (FOV) of $4.\!'33\times 4.\!'33$ with
a $1024\times 1024$ square pixel Rockwell HAWAII-I detector.  {\sl
WIRC} mounts at the telescope's f/3.3 prime focus. The instrument's
seeing-limited optical design, optimized for the JHK near-IR
atmospheric bands, includes a 4-element refractive collimator, two
7-position filter wheels that straddle a Lyot stop, and a 5-element
refractive f/3 camera. Typical seeing-limited point spread functions
are slightly oversampled with a $0.25''$/pixel plate scale at the
detector.

We took the observations on 22 March 2002 under good atmospheric
conditions but at high airmass ($1.9 - 2.2$).  Using Gaussian fits to
the stellar point sources within the FOV we determined FWHM$_{K_s}=0.9
\arcsec$ and FWHM$_{J}=1.0 \arcsec$.  The observing procedure was
controlled by a simple macro that commanded the telescope to 20 random
dither positions for better redundancy and enlarged field of view.  At
each dither position two and eight exposures were taken in the $J$ and
$K_s$ filters, respectively.  Because of the higher sky background the
integration times were shorter in $K_s$ (7.27\,s) than in $J$
(29.07\,s).  The resulting total integration times were 19.38 minutes
in both $J$ and $K_s$.

\begin{figure*}
\plotone{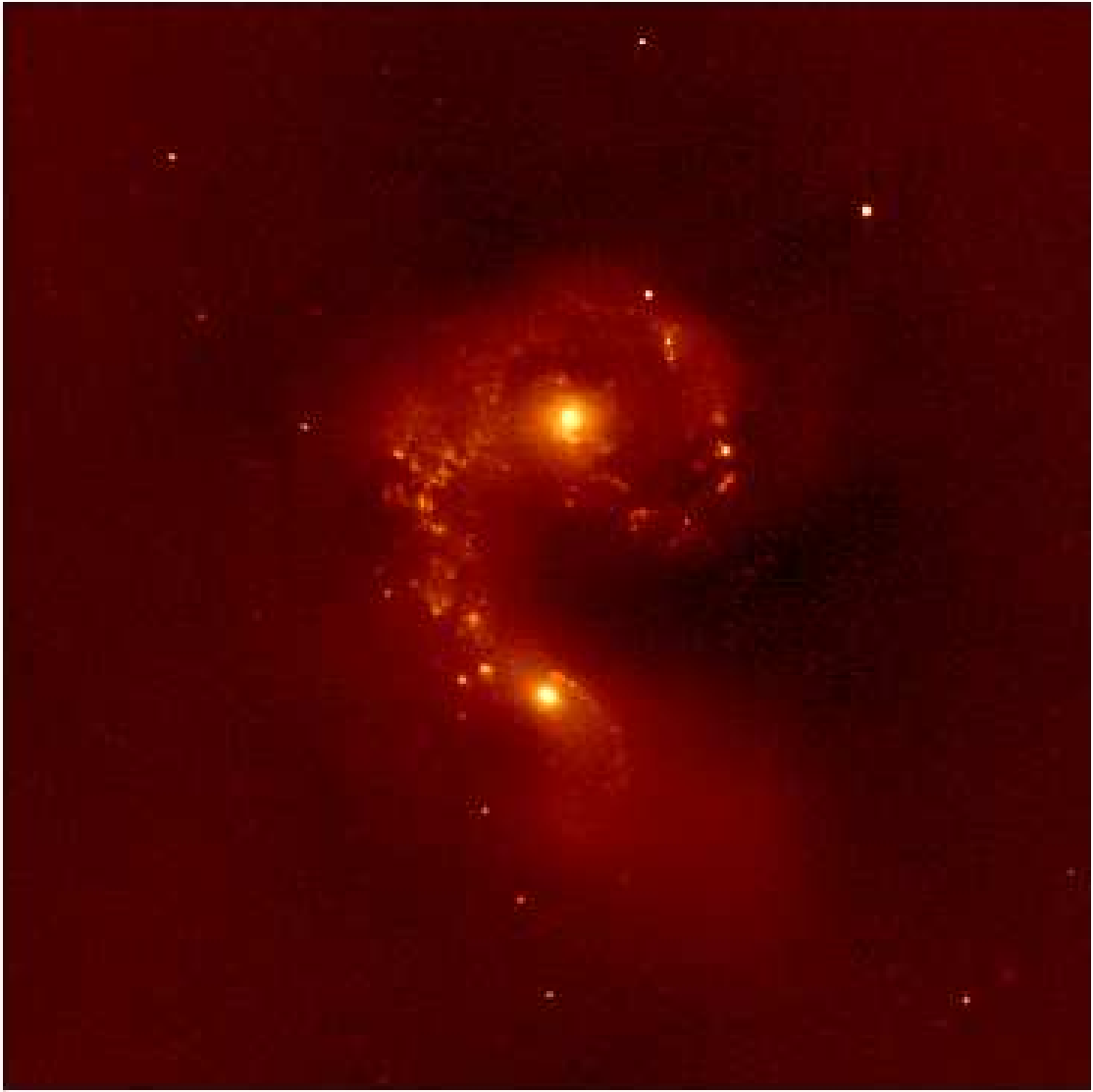}
\caption{{\sl WIRC} image of the Antennae galaxies NGC\,4038/39.  The
         image was taken in the $K_s$ filter and is $1024\times 1024$
         pixels ($4.\!'33\times 4.\!'33, 24.7\mbox{\,kpc}\times
         24.7\mbox{\,kpc}$) wide.  North is up and East to the
         left. The two brightest spots are the nuclei of NGC\,4038
         (top) and NGC\,4039 (bottom). The apparent alignment of point
         sources along a line from the overlap region to the south is
         a coincidence produced by foreground stars and also seen in
         optical images with larger FOV.  See Fig.~\ref{fignumbered}
         for individually labeled clusters.
\label{figantennaegal}}
\end{figure*}

%
%
\subsection{Data Reduction}
\label{secdatared}
We carried out the data reduction in {\sl IDL} using a self-written
procedure.  First, we averaged multiple exposures of each position and
subtracted the ``sky''.  Since no additional observations of a nearby,
blank sky had been taken, we computed the lowered (1/3 instead of 1/2)
median of the 20 dither positions.  The relative offsets between the
pointings were sufficiently large so that the filtered image looks
reasonably smooth with no noticeable residuals from the galaxies,
serving as a good ``sky'' image.  For diagnostic purposes we have also
rebinned the ``sky'' image to quantify a possible underlying gradient
from insufficiently subtracted diffuse emission of the galactic halos
and found that such a contribution must be less than 8\%, which
corresponds to a photometric error of 0.1 magnitudes.  However, this
value is only an upper limit since it may also include contributions from
a possible gradient in the sky background emission.

The next step was to apply the flat-field, which was computed from two
series of twilight exposures at high and low flux levels.  Five
difference images between high and low flux twilights have been
computed, inverted and normalized.  Finally, the median of those five
frames was calculated and renormalized.

The positional offset between the different images was derived from
cross-correlating the individual images relative to a reference frame
on an integer pixel grid.  Finally, the 20 aligned images were
combined via the median technique.

\begin{figure}[ht]
\plottwo{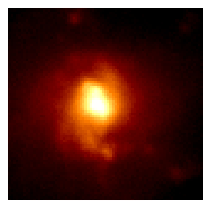}{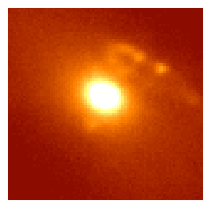}
\caption{The two galactic nuclei of the Antennae at $K_s$; left
         NGC\,4038, right NGC\,4039.  The nuclei have $J-K_s$ colors
         of 0.82 (NGC\,4038), and 0.95 (NGC\,4039).\label{fignuclei}}
\end{figure}

\begin{figure*}
\plottwo{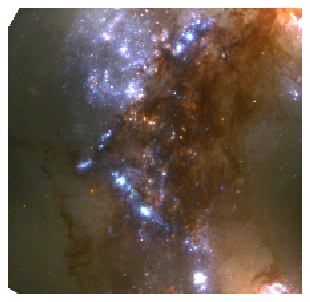}{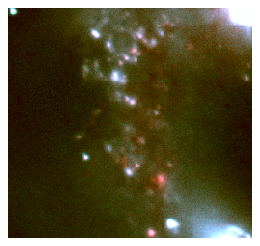}
\caption{A zoom into the Antennae overlap region.  Left: True-color
         HST/WFPC2 image in bands U+B (blue), V (green), and I (red)
         \citep{whi99}.  Right: False-color composite from WIRC $J$
         (blue \& green) and $K_s$ (red).  The two nuclei are still
         visible to the upper (NGC\,4038) and lower (NGC\,4039) right
         corner.  The two brightest red clusters are source \#148
         northeast of, and source \#157 east of the nucleus of NGC
         4039. \label{figoverlap}}
\end{figure*}

%
%
\subsection{Image Morphology}
The resulting $K_s$ image of the entire FOV is shown in
Figure~\ref{figantennaegal}.  (The $J$ band image shows similar
structure and is not shown here.)  The image is clearly dominated by
the two nuclei of the interacting galaxies.  In the Third Reference
Catalogue of Bright Galaxies \citep{vau91} NGC\,4038 is classified as
SB(s)m~pec ($B^0_T = 10.59$), while NGC\,4039 is classified as
SA(s)m~pec ($B^0_T = 10.69$).  Figure~\ref{fignuclei} shows the $K_s$
image of both nuclei in more detail.  While the nucleus of NGC\,4039
in the $K_s$ image looks very similar to the optical structure, the
nucleus of NGC\,4038 looks very different.  Most notably, the dust
lane in east-west direction has become almost invisible, replaced by
the prominent spiral arm structure of an Sb-type galaxy.

Besides the two nuclei, the star clusters in the western loop and the
southeastern overlap region are clearly visible.  A color zoom into
the overlap region is shown in Figure~\ref{figoverlap}, side-by-side
with the optical WFPC2 image from \citet{whi99}.

\begin{figure}
\plotone{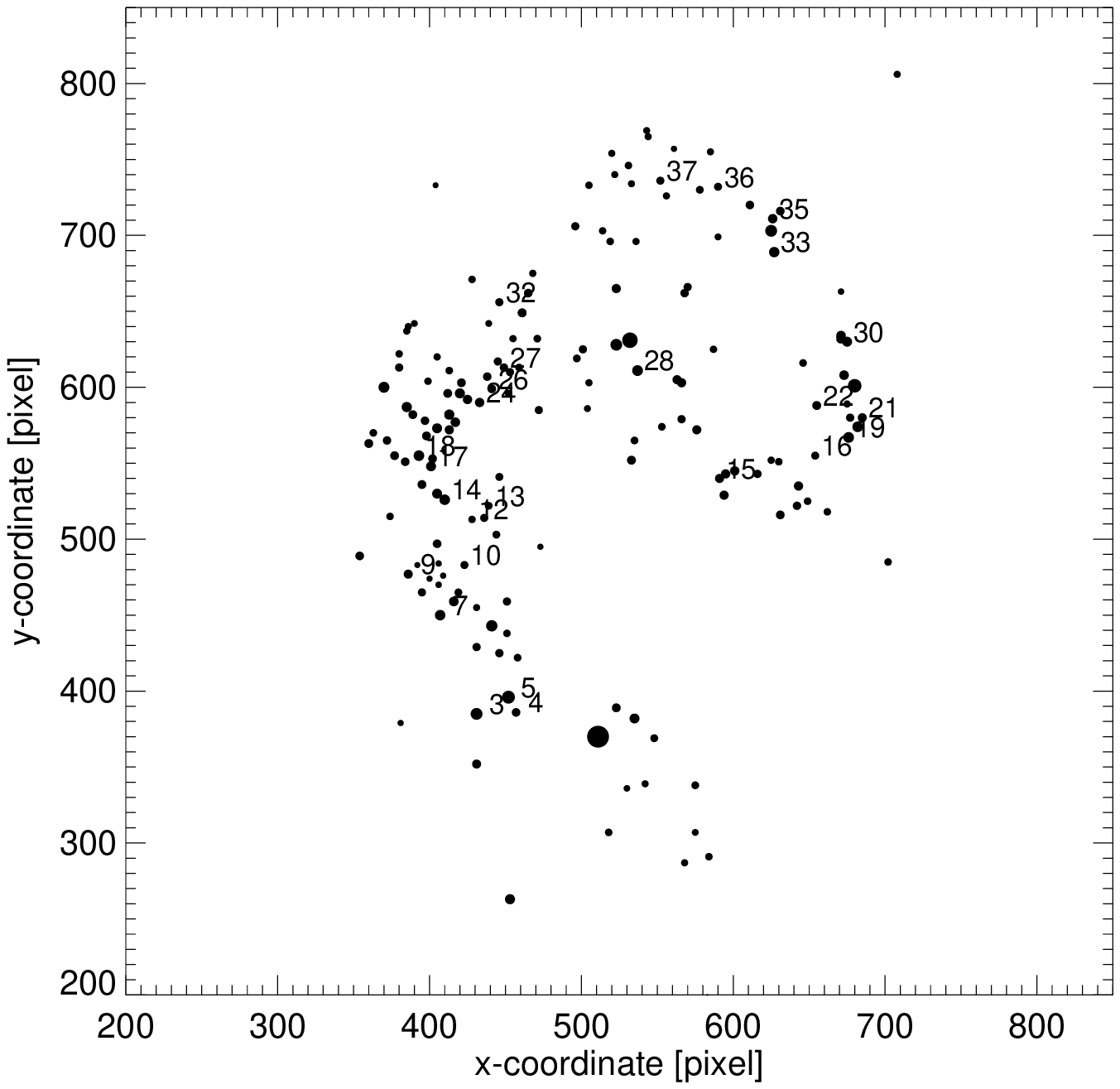}
\caption{The locations of the infrared clusters relative to the {\sl
         WIRC} pixel coordinates (x and y are approximately oriented
         east-west and north-south, respectively).  The numbers refer
         to the clusters with radio identifications also listed in
         Table~\ref{tabcombined}.  The numbers refer to clusters at
         their lower left, and the sizes of the dots scale with the
         cluster luminosities at $K_s$.\label{clusterlumi}}
\end{figure}

Figure~\ref{clusterlumi} illustrates the locations of all the near-IR
sources detected in the {\sl WIRC} images.  We discuss the infrared
sources in more detail in section~\ref{secdiscussion}.

\begin{figure*}
\epsscale{1.1}
\plotone{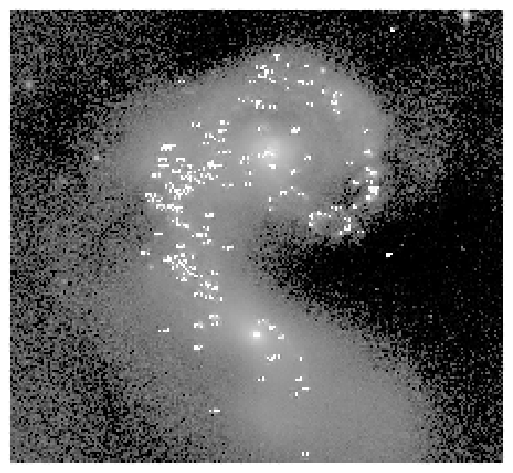}
\caption{K-band image of the Antennae with the WIRC source numbers
         overplotted to indicate the location of the clusters within
         the galaxy.  The nominal cluster position is at the center of
         each number. \label{fignumbered}}
\end{figure*}

%
%
\subsection{Cluster Identification}
\label{secclusterid}
Because of the varying shapes of the clusters and the gradients of the
underlying background the best method of cluster detection was by eye.
Small, compact ``smudges'' that could be distinguished from the
background and were detectable at both $J$ and $K_s$ were selected,
229 sources in total, which are labeled in Fig.~\ref{fignumbered}.

However, not all of these sources are good cluster candidates: eight
are known stars, one is a background quasar, and many more lie outside
the area with increased starburst activity.  They are therefore not
likely to be massive star clusters associated with the Antennae
galaxies.  Investigating their true nature is not the aim of this
paper; however, for completeness we list the 57 infrared sources,
which are likely not Antennae clusters, in Table~\ref{tabnonclus}.

\tabletypesize{\footnotesize}
\begin{deluxetable*}{rrrccccc}
\tablecaption{Sources within the WIRC images that are likely not star
              clusters that belong to the Antennae.  Source \#155 is a
              background quasar, and sources \#21, 27, 28, 30, 36, 64,
              159, 211 are known stars.  See Fig.~\ref{fignumbered} for
              their location within the galaxy.}
\tablehead{
\colhead{Id$_W$} & \colhead{RA (J2000)} & 
\colhead{Dec (J2000)} & \colhead{$J$} & \colhead{$\sigma (J)$} &
\colhead{$K_s$} & \colhead{$\sigma (K_s)$} & \colhead{($J-K_s$)}
}
\startdata
   1  & 12:01:44.64  & -18:53:55.04  &  16.94  &  0.02  &  16.20  &  0.01  &  0.74  \\
   2  & 12:01:44.57  & -18:52:35.51  &  19.39  &  0.04  &  17.36  &  0.02  &  2.03  \\
   3  & 12:01:44.81  & -18:54:24.56  &  18.06  &  0.03  &  17.94  &  0.07  &  0.12  \\
   4  & 12:01:44.80  & -18:53:26.43  &  19.34  &  0.03  &  17.84  &  0.02  &  1.50  \\
   5  & 12:01:45.04  & -18:53:04.93  &  18.79  &  0.02  &  18.40  &  0.04  &  0.39  \\
   6  & 12:01:45.71  & -18:54:20.70  &  18.69  &  0.04  &  16.99  &  0.02  &  1.70  \\
   7  & 12:01:45.79  & -18:53:46.00  &  19.24  &  0.04  &  17.88  &  0.03  &  1.36  \\
   8  & 12:01:45.75  & -18:51:34.71  &  17.90  &  0.03  &  17.45  &  0.04  &  0.45  \\
   9  & 12:01:46.47  & -18:54:25.30  &  14.90  &  0.01  &  15.03  &  0.01  & -0.13  \\
  10  & 12:01:46.79  & -18:53:29.36  &  19.29  &  0.14  &  19.17  &  0.14  &  0.12  \\
  11  & 12:01:46.80  & -18:53:26.56  &  19.88  &  0.21  &  20.55  &  0.99  & -0.67  \\
  12  & 12:01:46.91  & -18:54:30.77  &  18.33  &  0.03  &  17.48  &  0.10  &  0.85  \\
  13  & 12:01:46.94  & -18:54:00.17  &  20.13  &  0.08  &  20.12  &  0.19  &  0.01  \\
  15  & 12:01:47.47  & -18:54:40.28  &  19.08  &  0.09  &  19.01  &  0.11  &  0.07  \\
  16  & 12:01:47.61  & -18:54:24.68  &  19.62  &  0.02  &  17.66  &  0.03  &  1.96  \\
  17  & 12:01:47.33  & -18:50:59.02  &  17.98  &  0.02  &  18.19  &  0.09  & -0.21  \\
  18  & 12:01:47.44  & -18:51:20.14  &  18.81  &  0.10  &  19.29  &  0.11  & -0.48  \\
  19  & 12:01:47.73  & -18:54:19.55  &  19.11  &  0.07  &  17.29  &  0.02  &  1.82  \\
  21  & 12:01:47.90  & -18:51:15.66  &  13.07  &  0.01  &  12.77  &  0.00  &  0.30  \\
  22  & 12:01:48.13  & -18:52:40.49  &  18.40  &  0.01  &  17.31  &  0.03  &  1.09  \\
  23  & 12:01:48.17  & -18:51:09.71  &  19.28  &  0.04  &  19.48  &  0.34  & -0.20  \\
  24  & 12:01:48.56  & -18:53:41.55  &  19.14  &  0.03  &  18.17  &  0.06  &  0.97  \\
  25  & 12:01:48.56  & -18:53:25.99  &  19.32  &  0.04  &  19.04  &  0.16  &  0.28  \\
  26  & 12:01:48.97  & -18:53:04.70  &  18.12  &  0.02  &  17.31  &  0.02  &  0.81  \\
  27  & 12:01:48.97  & -18:53:04.70  &  18.20  &  0.01  &  17.31  &  0.02  &  0.89  \\
  28  & 12:01:49.03  & -18:53:22.02  &  19.33  &  0.06  &  18.59  &  0.06  &  0.74  \\
  29  & 12:01:48.88  & -18:50:38.89  &  19.60  &  0.09  &  18.96  &  0.08  &  0.64  \\
  30  & 12:01:49.78  & -18:51:20.42  &  17.85  &  0.03  &  17.77  &  0.04  &  0.08  \\
  33  & 12:01:50.43  & -18:54:15.12  &  17.69  &  0.01  &  18.14  &  0.05  & -0.45  \\
  36  & 12:01:50.38  & -18:52:53.80  &  18.71  &  0.03  &  18.64  &  0.11  &  0.07  \\
  64  & 12:01:51.63  & -18:51:34.61  &  14.84  &  0.01  &  14.24  &  0.01  &  0.60  \\
 113  & 12:01:53.58  & -18:54:21.76  &  15.94  &  0.01  &  15.41  &  0.01  &  0.53  \\
 122  & 12:01:54.06  & -18:53:58.92  &  16.28  &  0.01  &  15.50  &  0.01  &  0.78  \\
 143  & 12:01:54.78  & -18:54:09.90  &  19.65  &  0.05  &  18.00  &  0.02  &  1.65  \\
 155  & 12:01:54.97  & -18:53:14.51  &  17.55  &  0.02  &  16.16  &  0.01  &  1.39  \\
 159  & 12:01:54.87  & -18:51:47.35  &  18.01  &  0.01  &  17.04  &  0.02  &  0.97  \\
 211  & 12:01:56.20  & -18:52:44.80  &  16.49  &  0.02  &  16.37  &  0.01  &  0.12  \\
 215  & 12:01:57.36  & -18:51:36.87  &  19.08  &  0.03  &  17.11  &  0.02  &  1.97  \\
 216  & 12:01:57.56  & -18:52:04.59  &  15.56  &  0.01  &  15.16  &  0.01  &  0.40  \\
 217  & 12:01:58.22  & -18:51:17.21  &  18.31  &  0.03  &  17.90  &  0.06  &  0.41  \\
 218  & 12:01:58.44  & -18:52:49.44  &  17.64  &  0.02  &  16.90  &  0.04  &  0.74  \\
 219  & 12:01:58.44  & -18:52:49.44  &  17.56  &  0.01  &  16.90  &  0.04  &  0.66  \\
 220  & 12:01:58.41  & -18:52:13.50  &  18.59  &  0.02  &  17.05  &  0.02  &  1.54  \\
 221  & 12:01:58.74  & -18:52:53.68  &  18.89  &  0.03  &  18.53  &  0.03  &  0.36  \\
 222  & 12:01:58.93  & -18:54:13.93  &  19.15  &  0.04  &  18.75  &  0.06  &  0.40  \\
 223  & 12:01:59.26  & -18:51:37.28  &  17.78  &  0.01  &  15.97  &  0.01  &  1.81  \\
 224  & 12:01:59.91  & -18:54:32.23  &  18.82  &  0.02  &  18.20  &  0.03  &  0.62  \\
 225  & 12:01:59.72  & -18:50:58.89  &  15.86  &  0.01  &  14.91  &  0.01  &  0.95  \\
 226  & 12:02:00.11  & -18:54:32.68  &  19.23  &  0.03  &  18.48  &  0.07  &  0.75  \\
 227  & 12:02:00.26  & -18:54:34.41  &  18.79  &  0.02  &  18.42  &  0.11  &  0.37  \\
 228  & 12:02:00.00  & -18:50:27.18  &  18.94  &  0.02  &  18.71  &  0.07  &  0.23  \\
 229  & 12:02:00.37  & -18:51:07.60  &  18.56  &  0.03  &  17.74  &  0.04  &  0.82  \\
 230  & 12:02:00.39  & -18:51:11.42  &  19.10  &  0.03  &  20.09  &  0.14  & -0.99  \\
 231  & 12:02:00.92  & -18:54:22.48  &  18.85  &  0.12  &  19.49  &  0.05  & -0.64  \\
 232  & 12:02:01.47  & -18:51:56.97  &  19.42  &  0.04  &  20.21  &  0.24  & -0.79  \\
 233  & 12:02:01.78  & -18:51:31.37  &  20.11  &  0.05  &  17.45  &  0.04  &  2.66  \\
 234  & 12:02:01.87  & -18:51:28.54  &  19.56  &  0.03  &  17.17  &  0.03  &  2.39  \\
\enddata
\label{tabnonclus}
\end{deluxetable*}
\tabletypesize{\normalsize}

Subtracting these contaminating sources we found a total of 172 good
Antennae cluster candidates with a signal-to-noise ratio $S/N \ge 5$
in both filter bands.  For that signal-to-noise ratio we derive
position-dependent completeness limits of $J \approx 19.0$ and $K_s
\approx 19.4$.  The reddest clusters with $J-K_s > 1$ are listed in
Table~\ref{tabreddest}.

\tabletypesize{\footnotesize}
\begin{deluxetable*}{rrrccccc}
\tablecaption{The reddest clusters ($J-K_s > 1$) within the WIRC image
              of the Antennae.  The nuclei of NGC\,4038 (\#102) and
              NGC\,4039 (\#114) have $J-K_s$ colors of 0.82, and 0.95,
              respectively, and are therefore not included in the
              table. Source \#148 coincides with the peak of
              luminosity in the mid-IR and sub-mm maps.}
\tablehead{
\colhead{Id$_W$\tablenotemark{a}} & \colhead{RA (J2000)} & 
\colhead{Dec (J2000)} & \colhead{$J$} & \colhead{$\sigma (J)$} &
\colhead{$K_s$} & \colhead{$\sigma (K_s)$} & \colhead{($J-K_s$)}
}
\startdata
 135  & 12:01:54.48  & -18:52:08.89  &  20.30  &  0.45  &  16.84  &  0.03  &  3.46  \\
 190  & 12:01:55.58  & -18:52:45.51  &  19.16  &  0.06  &  16.79  &  0.03  &  2.37  \\
 125  & 12:01:54.14  & -18:52:03.39  &  19.51  &  0.08  &  17.21  &  0.02  &  2.30  \\
 139  & 12:01:54.57  & -18:52:47.36  &  18.67  &  0.03  &  16.72  &  0.02  &  1.95  \\
 148  & 12:01:54.76  & -18:52:51.38  &  16.76  &  0.02  &  14.88  &  0.04  &  1.88  \\
 157  & 12:01:54.96  & -18:53:06.10  &  16.52  &  0.01  &  14.66  &  0.01  &  1.86  \\
 111  & 12:01:53.41  & -18:53:26.49  &  19.20  &  0.04  &  17.36  &  0.02  &  1.84  \\
 128  & 12:01:54.32  & -18:51:59.00  &  18.21  &  0.02  &  16.38  &  0.01  &  1.83  \\
 141  & 12:01:54.55  & -18:52:08.10  &  18.38  &  0.89  &  16.63  &  0.01  &  1.75  \\
 144  & 12:01:54.59  & -18:51:57.13  &  18.50  &  0.03  &  16.81  &  0.02  &  1.69  \\
 124  & 12:01:54.14  & -18:52:15.37  &  18.51  &  0.02  &  16.85  &  0.01  &  1.66  \\
 129  & 12:01:54.32  & -18:51:59.00  &  17.81  &  0.01  &  16.28  &  0.03  &  1.53  \\
  84  & 12:01:52.37  & -18:51:56.30  &  18.06  &  0.07  &  16.56  &  0.02  &  1.50  \\
 169  & 12:01:55.21  & -18:52:08.40  &  18.80  &  0.07  &  17.31  &  0.02  &  1.49  \\
 165  & 12:01:55.08  & -18:52:12.27  &  16.99  &  0.03  &  15.50  &  0.01  &  1.49  \\
 167  & 12:01:55.14  & -18:52:17.09  &  17.25  &  0.03  &  15.77  &  0.01  &  1.48  \\
 126  & 12:01:54.18  & -18:51:52.41  &  19.08  &  0.03  &  17.60  &  0.03  &  1.48  \\
 116  & 12:01:53.54  & -18:52:10.98  &  19.15  &  0.52  &  17.71  &  0.03  &  1.44  \\
 182  & 12:01:55.36  & -18:52:18.04  &  16.98  &  0.10  &  15.55  &  0.03  &  1.43  \\
  32  & 12:01:50.01  & -18:52:42.19  &  18.80  &  0.04  &  17.40  &  0.02  &  1.40  \\
 168  & 12:01:55.21  & -18:52:47.16  &  17.10  &  0.07  &  15.71  &  0.02  &  1.39  \\
  46  & 12:01:50.43  & -18:50:40.43  &  19.33  &  0.03  &  18.00  &  0.04  &  1.33  \\
 172  & 12:01:55.23  & -18:52:12.22  &  17.61  &  0.05  &  16.28  &  0.03  &  1.33  \\
  52  & 12:01:50.97  & -18:52:08.48  &  18.59  &  0.02  &  17.27  &  0.03  &  1.32  \\
  63  & 12:01:51.57  & -18:51:41.77  &  17.78  &  0.01  &  16.49  &  0.02  &  1.29  \\
  77  & 12:01:52.27  & -18:53:54.13  &  18.73  &  0.03  &  17.48  &  0.02  &  1.25  \\
  75  & 12:01:52.22  & -18:53:30.95  &  18.62  &  0.03  &  17.37  &  0.02  &  1.25  \\
  82  & 12:01:52.34  & -18:51:55.30  &  18.01  &  0.02  &  16.77  &  0.01  &  1.24  \\
 206  & 12:01:55.95  & -18:52:32.64  &  18.66  &  0.05  &  17.42  &  0.02  &  1.24  \\
 106  & 12:01:53.20  & -18:52:04.71  &  15.95  &  0.01  &  14.71  &  0.15  &  1.24  \\
  71  & 12:01:51.95  & -18:51:38.59  &  18.15  &  0.05  &  16.92  &  0.02  &  1.23  \\
 188  & 12:01:55.49  & -18:52:19.27  &  17.52  &  0.01  &  16.29  &  0.02  &  1.23  \\
 170  & 12:01:55.22  & -18:52:15.79  &  16.63  &  0.01  &  15.40  &  0.01  &  1.23  \\
  95  & 12:01:52.96  & -18:53:18.47  &  19.08  &  0.02  &  17.86  &  0.02  &  1.22  \\
  98  & 12:01:53.01  & -18:52:20.83  &  18.36  &  0.07  &  17.16  &  0.02  &  1.20  \\
 187  & 12:01:55.46  & -18:52:10.11  &  18.64  &  0.21  &  17.45  &  0.03  &  1.19  \\
 145  & 12:01:54.67  & -18:52:56.00  &  17.86  &  0.02  &  16.67  &  0.02  &  1.19  \\
 130  & 12:01:54.37  & -18:52:08.16  &  18.36  &  0.04  &  17.17  &  0.02  &  1.19  \\
 118  & 12:01:53.60  & -18:52:05.35  &  17.93  &  0.26  &  16.74  &  0.15  &  1.19  \\
  78  & 12:01:52.17  & -18:51:39.03  &  18.29  &  0.02  &  17.11  &  0.08  &  1.18  \\
 117  & 12:01:53.56  & -18:52:15.30  &  19.14  &  0.03  &  17.96  &  0.05  &  1.18  \\
 142  & 12:01:54.63  & -18:52:26.43  &  18.14  &  0.07  &  16.97  &  0.04  &  1.17  \\
 183  & 12:01:55.32  & -18:51:37.26  &  20.54  &  0.15  &  19.38  &  0.22  &  1.16  \\
 171  & 12:01:55.22  & -18:52:18.34  &  17.17  &  0.11  &  16.03  &  0.06  &  1.14  \\
  92  & 12:01:52.84  & -18:53:10.86  &  18.17  &  0.02  &  17.04  &  0.14  &  1.13  \\
  97  & 12:01:52.94  & -18:51:47.45  &  18.79  &  0.05  &  17.68  &  0.02  &  1.11  \\
 151  & 12:01:54.77  & -18:52:31.24  &  18.29  &  0.05  &  17.18  &  0.03  &  1.11  \\
  94  & 12:01:52.79  & -18:51:28.89  &  19.02  &  0.09  &  17.92  &  0.03  &  1.10  \\
 156  & 12:01:54.94  & -18:52:54.89  &  17.83  &  0.02  &  16.73  &  0.03  &  1.10  \\
 153  & 12:01:54.82  & -18:52:33.26  &  17.97  &  0.02  &  16.88  &  0.02  &  1.09  \\
 185  & 12:01:55.45  & -18:52:24.39  &  16.63  &  0.06  &  15.55  &  0.05  &  1.08  \\
 179  & 12:01:55.35  & -18:52:06.06  &  18.63  &  0.02  &  17.56  &  0.03  &  1.07  \\
 115  & 12:01:53.49  & -18:51:37.84  &  18.56  &  0.01  &  17.50  &  0.02  &  1.06  \\
  91  & 12:01:52.64  & -18:51:37.35  &  17.93  &  0.03  &  16.88  &  0.02  &  1.05  \\
 207  & 12:01:55.97  & -18:52:19.89  &  17.61  &  0.03  &  16.58  &  0.02  &  1.03  \\
 110  & 12:01:53.25  & -18:51:47.36  &  18.39  &  0.02  &  17.36  &  0.02  &  1.03  \\
 173  & 12:01:55.29  & -18:52:30.05  &  16.25  &  0.05  &  15.23  &  0.03  &  1.02  \\
  90  & 12:01:52.68  & -18:52:18.64  &  18.24  &  0.17  &  17.23  &  0.04  &  1.01  \\
  58  & 12:01:51.29  & -18:51:49.77  &  16.34  &  0.05  &  15.33  &  0.01  &  1.01  \\
 177  & 12:01:55.38  & -18:52:40.73  &  19.67  &  0.16  &  18.67  &  0.24  &  1.00  \\
\enddata
\label{tabreddest}
\tablenotetext{a}{Identification number from the {\sl WIRC} image.}
\end{deluxetable*}
\tabletypesize{\normalsize}

We determined the source positions via cross-correlation between {\sl
WIRC} and 2MASS \citep{skr05}.  We identified six bright, compact IR
sources common to both {\sl WIRC} and 2MASS images.  Using a least
squares fit to a two-dimensional linear matching function, we
associated {\sl WIRC} x,y pixel centroids to the 2MASS right ascension
and declination, yielding a positional uncertainty of $\approx 0.2''$.
Considering the systematic offsets between the 2MASS, {\sl HST} and
{\sl Chandra} coordinate systems in the order of $1''$, this
uncertainty seems negligible.  Figure~\ref{clusterlumi} illustrates
the location of the detected infrared clusters.

%
%
\subsection{Photometry}
\label{secphotometry}
To derive the photometric fluxes we used a self-written {\sl IDL}
procedure for simple aperture photometry with a 5-pixel radius at
$K_s$, and a 6-pixel radius at $J$, corresponding to $\sim 3 \sigma$
of the Gaussian PSF.  In order to estimate the uncertainties in
background subtraction, we measured the background in two separate
annuli around each source: one from 9 to 12 pixels and another from 12
to 15 pixels. Due to the high concentration of clusters, we employed
the use of sky background arcs instead of annuli for some sources.
These were defined by a position angle and opening angle with respect
to the source center.  All radii were kept constant to ensure
consistency.  In addition, nearby bright sources could shift the
computed central peak position by as much as a pixel or two.  If the
centroid position determined for a given source differed by more than
one pixel from the apparent peak due to such contamination, we forced
the center of the photometric apertures to be at the apparent peak
position.  For both annular regions, we calculated the mean and median
backgrounds per pixel.  

Multiplying the backgrounds by the area of the central aperture, these
values were subtracted from the flux measurement of the central
aperture to yield four flux values for the source in terms of DN.
Averaging the four values provided us with a flux value for each
cluster.  We did not apply any aperture correction factors since these
are negligible for our seeing-limited PSFs.  We computed errors by
considering both variations in sky background, $\sigma_{sky}$, and
Poisson noise, $\sigma_{adu}$.  We computed $\sigma_{sky}$ by taking
the standard deviation of the four measured flux values.  We then
calculated the expected Poisson noise by scaling DN to $e^-$ using the
known gain of {\sl WIRC} \citep{wil03} and taking the square root of
this value.  We added both terms in quadrature for the total
photometric errors that are listed in Tables~\ref{tabreddest} and
\ref{tabcombined}.

Finally, we calibrated the derived source fluxes with the bright 2MASS
star `2MASS 12014790-1851156' at $12^h 01^m 47.90^s$, $-18^d 51^m
15.7^s$ which is listed in the 2MASS database with $J=13.065$ and
$K_s=12.771$, and was in the {\sl WIRC} FOV during the Antennae
observations.

\begin{deluxetable*}{rrrrrrrccccccc}
\tabletypesize{\scriptsize}
\tablecaption{The 27 Antennae clusters which have been detected by all
              three {\sl WIRC}, {\sl HST} and the VLA.}
\tablewidth{0pt}
\tablehead{
\colhead{Id$_H$\tablenotemark{a}} & 
\colhead{Id$_W$\tablenotemark{b}} & 
\colhead{RA (J2000)} & \colhead{Dec (J2000)} & 
\colhead{Diff\tablenotemark{c}} &
\colhead{Age\tablenotemark{d}} & 
\colhead{Flux\tablenotemark{e}} & 
\colhead{$M_V$} & \colhead{$m_V$} & \colhead{$m_I$} & 
\colhead{$m_J$} & \colhead{$\sigma (J)$} &
\colhead{$m_K$} & \colhead{$\sigma (K)$}
}
\startdata
   21  &   34  & 12:01:50.28  & -18:52:17.88  &  0.96  &  4.50  &    49  & -11.37  &  20.04  &  20.19  &
16.72  &  0.02  &  16.17  &  0.01  \\
  19  &   39  & 12:01:50.45  & -18:52:21.14  &  0.15  &  6.40  &   217  & -12.12  &  19.63  &  19.36  &
15.86  &  0.03  &  15.12  &  0.02  \\
  30  &   40  & 12:01:50.44  & -18:52:05.08  &  0.51  &  5.20  &    80  & -12.67  &  19.09  &  18.97  &
16.18  &  0.02  &  15.57  &  0.03  \\
  22  &   48  & 12:01:50.82  & -18:52:15.67  &  0.55  &  4.00  &   102  & -10.57  &  21.00  &  20.78  &
17.09  &  0.01  &  16.21  &  0.01  \\
  16  &   49  & 12:01:50.85  & -18:52:24.07  &  0.68  &  5.20  &    32  & -10.60  &  21.59  &  21.33  &
17.70  &  0.01  &  16.94  &  0.03  \\
  33  &   58  & 12:01:51.29  & -18:51:49.77  &  0.50  &  7.00  &    59  & -13.33  &  20.03  &  18.88  &
16.34  &  0.05  &  15.33  &  0.01  \\
  35  &   59  & 12:01:51.30  & -18:51:44.15  &  1.10  &  7.70  &    43  & -11.19  &  20.82  &  20.19  &
16.34  &  0.06  &  15.86  &  0.01  \\
  15  &   70  & 12:01:52.00  & -18:52:27.53  &  0.75  &  6.00  &   113  & -12.05  &  19.66  &  19.24  &
16.82  &  0.01  &  16.10  &  0.01  \\
  36  &   71  & 12:01:51.95  & -18:51:38.59  &  0.01  &  2.00  &   191  & -12.62  &  21.90  &  21.04  &
18.15  &  0.05  &  16.92  &  0.02  \\
  37  &   91  & 12:01:52.64  & -18:51:37.35  &  0.29  &  2.00  &    71  & -11.67  &  19.75  &  19.63  &
17.93  &  0.03  &  16.88  &  0.02  \\
  28  &   96  & 12:01:52.95  & -18:52:09.12  &  1.05  &  2.00  &   145  & -13.11  &  18.51  &  18.44  &
15.77  &  0.01  &  15.13  &  0.06  \\
   4  &  132  & 12:01:54.49  & -18:53:06.00  &  0.74  &  3.70  &   241  & -11.97  &  21.52  &  20.69  &
16.71  &  0.08  &  16.45  &  0.07  \\
   5  &  136  & 12:01:54.58  & -18:53:03.42  &  0.02  &  3.80  &  2257  & -14.50  &  19.07  &  18.40  &
15.05  &  0.03  &  14.27  &  0.02  \\
  27  &  141  & 12:01:54.55  & -18:52:08.10  &  0.52  &  5.00  &    88  & -13.21  &  19.20  &  19.26  &
18.38  &  0.89  &  16.63  &  0.01  \\
  32  &  144  & 12:01:54.59  & -18:51:57.13  &  0.45  &  2.00  &   493  & -13.16  &  21.26  &  20.36  &
18.50  &  0.03  &  16.81  &  0.02  \\
  26  &  149  & 12:01:54.70  & -18:52:11.62  &  0.60  &  4.00  &    74  & -12.55  &  18.87  &  18.84  &
17.17  &  0.06  &  16.35  &  0.01  \\
  13  &  151  & 12:01:54.77  & -18:52:31.24  &  0.63  &  2.00  &   341  & -13.93  &  22.64  &  20.80  &
18.29  &  0.05  &  17.18  &  0.03  \\
  24  &  154  & 12:01:54.85  & -18:52:13.87  &  0.50  &  8.40  &   120  & -11.85  &  20.74  &  19.75  &
16.60  &  0.03  &  15.92  &  0.01  \\
   3  &  157  & 12:01:54.96  & -18:53:06.10  &  0.00  &  2.00  &  5161  & -15.51  &  23.52  &  20.60  &
16.52  &  0.01  &  14.66  &  0.01  \\
  12  &  160  & 12:01:54.97  & -18:52:33.47  &  1.15  &  4.80  &    61  & -12.06  &  21.87  &  21.08  &
18.11  &  0.02  &  17.47  &  0.04  \\
  10  &  163  & 12:01:55.07  & -18:52:41.08  &  1.10  &  2.00  &   616  & -12.32  &  22.54  &  21.30  &
17.81  &  0.01  &  16.89  &  0.02  \\
  14  &  173  & 12:01:55.29  & -18:52:30.05  &  0.79  &  5.00  &    76  & -12.98  &  22.34  &  20.78  &
16.25  &  0.05  &  15.23  &  0.03  \\
   7  &  176  & 12:01:55.37  & -18:52:49.40  &  0.47  &  2.00  &  1215  & -13.62  &  19.06  &  18.92  &
 16.21  &  0.02  &  15.27  &  0.04  \\
  17  &  185  & 12:01:55.45  & -18:52:24.39  &  0.57  &  2.50  &   262  & -12.19  &  20.58  &  20.06  &
 16.63  &  0.06  &  15.55  &  0.05  \\
  18  &  192  & 12:01:55.59  & -18:52:22.56  &  1.17  &  7.40  &   217  & -11.92  &  20.31  &  19.65  &
 15.86  &  0.07  &  15.21  &  0.04  \\
   9  &  198  & 12:01:55.74  & -18:52:42.40  &  0.58  &  3.00  &   243  & -12.73  &  18.70  &  18.57  &
 16.70  &  0.02  &  16.10  &  0.02  \\
\enddata
\label{tabcombined}
\tablenotetext{a}{Identification number from the {\sl HST} image \citep{whi02}.}
\tablenotetext{b}{Identification number from the {\sl WIRC} image.}
\tablenotetext{c}{Radial difference between {\sl HST} and {\sl WIRC} position 
                  in arcseconds.}
\tablenotetext{d}{Age in million years according to \citet{whi02}.}
\tablenotetext{e}{Flux density at 6\,cm in mJy from \citet{nef00}.}
\end{deluxetable*}
\tabletypesize{\normalsize}


\section{Discussion of the Cluster Properties}
\label{secdiscussion}


\subsection{The Reddest Clusters}
\label{reddest}
\begin{figure}
\plotone{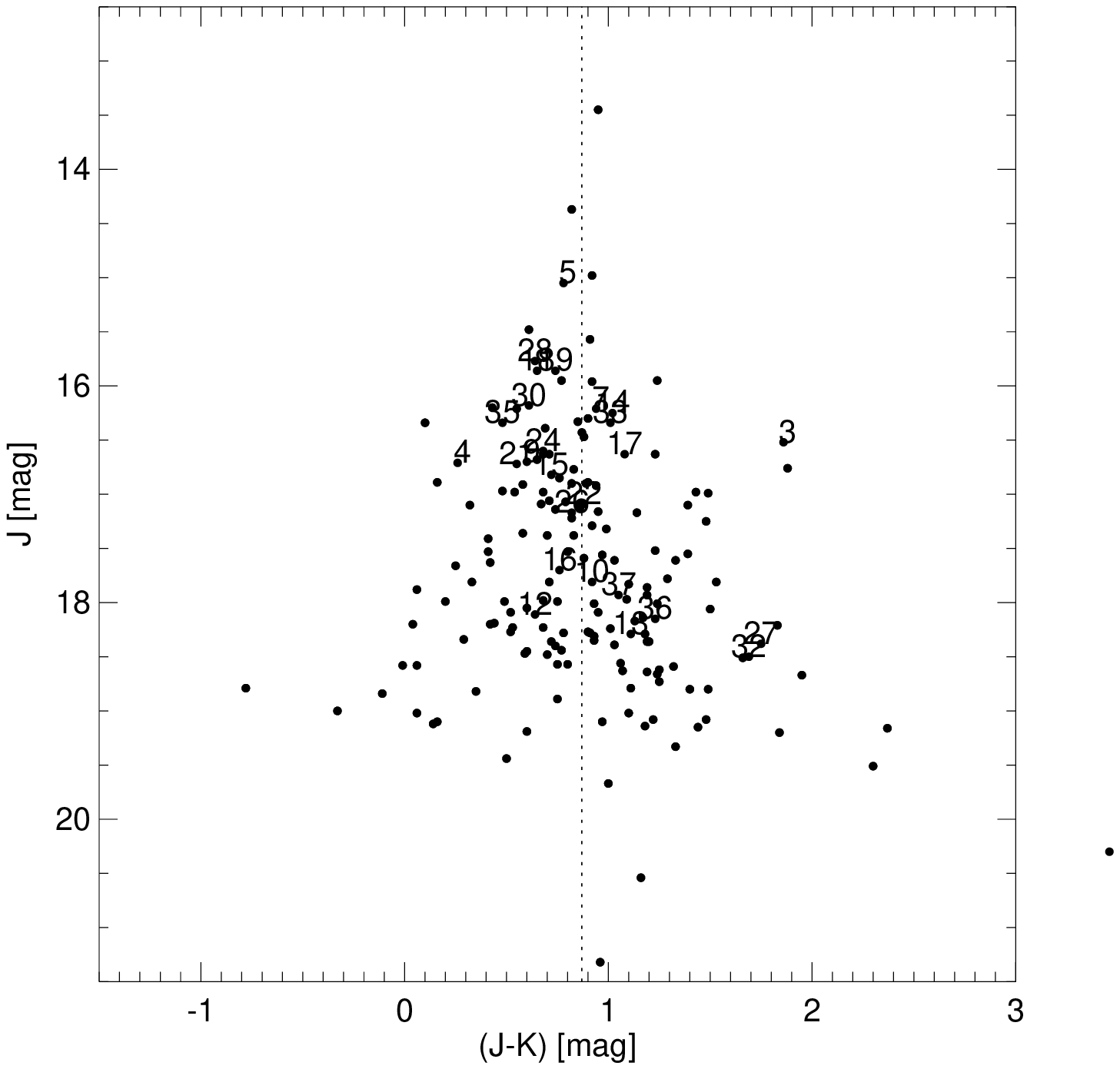}
\caption{$J$ vs.($J-K_s$) color-magnitude diagram for the infrared
         clusters.  The numbers refer to the clusters with radio
         identification in Table~\ref{tabcombined}.  The median color
         of $J-K_s = 0.87$ is indicated by a dashed vertical line; the
         $1-\sigma$ scatter is 0.5 magnitudes.  The two data point
         with the brightest $J$ magnitudes are the southern and norther
         nuclei.
\label{clustercmd}}
\end{figure}

A side-by-side comparison between the $UBVI$ WFPC2 and the WIRC $JK$
images (Figure~\ref{figoverlap}) reveals distinct differences in the
highly obscured overlap region, most notibly the bright cluster \#148.
If we simply assume that the colors are due to foreground reddening in
the overlap region, and not intrinsic to a cluster,
Figure~\ref{figoverlap} would represent an impressive 3-dimensional
extinction map of the region, with the bluest clusters being located
closest (or in voids) relative to the observer.  Fig.~\ref{clustercmd}
shows the $J$ vs.($J-K_s$) color-magnitude diagram for all infrared
cluster sources, including the ones in Table~\ref{tabcombined}.

\citet{kas03}, using ground-based observations at $1.\!''5$ seeing,
have found five red clusters in the overlap region with $A_V\geq 3$,
which are too faint in the optical to show up in their $B$- or
$V$-band images.  Interestingly, $A_V\sim 3$ corresponds to only
$(J-K_s)\sim 0.5$ \citep{rie85}, which is by far not near the blue end
of the color distribution in Fig.~\ref{clustercmd}.  Since higher
spatial resolution is expected to yield a larger range in extinction,
the values found by \citet{kas03} under $1.\!''5$ may just represent a
lower limit.  For comparison, extinction estimates from optical
observations at the diffraction limit have been discussed by
\citet{whi02}.

The observed color spread in Fig.~\ref{clustercmd} can be caused by
two effects, which are difficult to disentangle: extinction and
stellar evolution.  We have used the starburst modelling package {\it
Starburst99} by \citet{lei99} to estimate the mean $(J-K_s)$ color of
a young cluster and its evolution with time.  For solar metallicity,
an upper mass cutoff of $100 M_{\odot}$, a Salpeter initial mass
function \citep{sal55}, and an instantaneous burst, the {\it
Starburst99}
website\footnote{http://www.stsci.edu/science/starburst99/} provides
the following information:
A 1~Myr old cluster has $(J-K)\sim 0.65$, which slowly decreases with
time: at an age of $3-4$~Myr the cluster reaches its blue maximum of
$(J-K)\sim 0.23$, which is mainly due to the evolution of its most
massive members into Wolf-Rayet stars.  The mean color over its first
$6-7$~Myr is $(J-K)\sim 0.5$.  Thereafter, red supergiants appear and
the cluster reaches its reddest color of $(J-K)\sim 1.1$ at about
10~Myr. Much older clusters remain at $(J-K)\sim 0.6$ with little
color evolution from $10^{7.5}-10^9$~years.

If we adopt the ages derived by \citet{whi02} for the clusters in
Table~\ref{tabcombined} their mean intrinsic near-IR color is hence
$(J-K)\sim 0.5$.  The median color of our clusters, illustrated in
Fig.~\ref{clustercmd}, is 0.37~mag above the theoretical average,
corresponding to $A_V=2.2$\,mag \citep{rie85}.  In other words, the
clusters listed in Table~\ref{tabcombined} experience about two
magnitudes of optical extinction, on average.  Clusters that are bluer
than our median value of 0.87 may either suffer from less extinction
or have intrinsically bluer colors. However, clusters which are redder
than $(J-K)= 1.1$ (the reddest {\it Starburst99} color) are likely
to be heavily extincted.  Clusters with $(J-K_s)\sim 2.3$ thus have
$A_V\ge 7$.  If we assume young ages of less than 7~Myr
(Tab.~\ref{tabcombined}) the theoretically reddest color is
$(J-K_s)\sim 0.65$ and the individual reddening may be as high as 9.6
mag.  This value is in excellent agreement with \citet{gil00} who
derived a screen extinction of $A_V\sim 9-10$ toward their reddest
cluster \#2.

\begin{figure}
\plotone{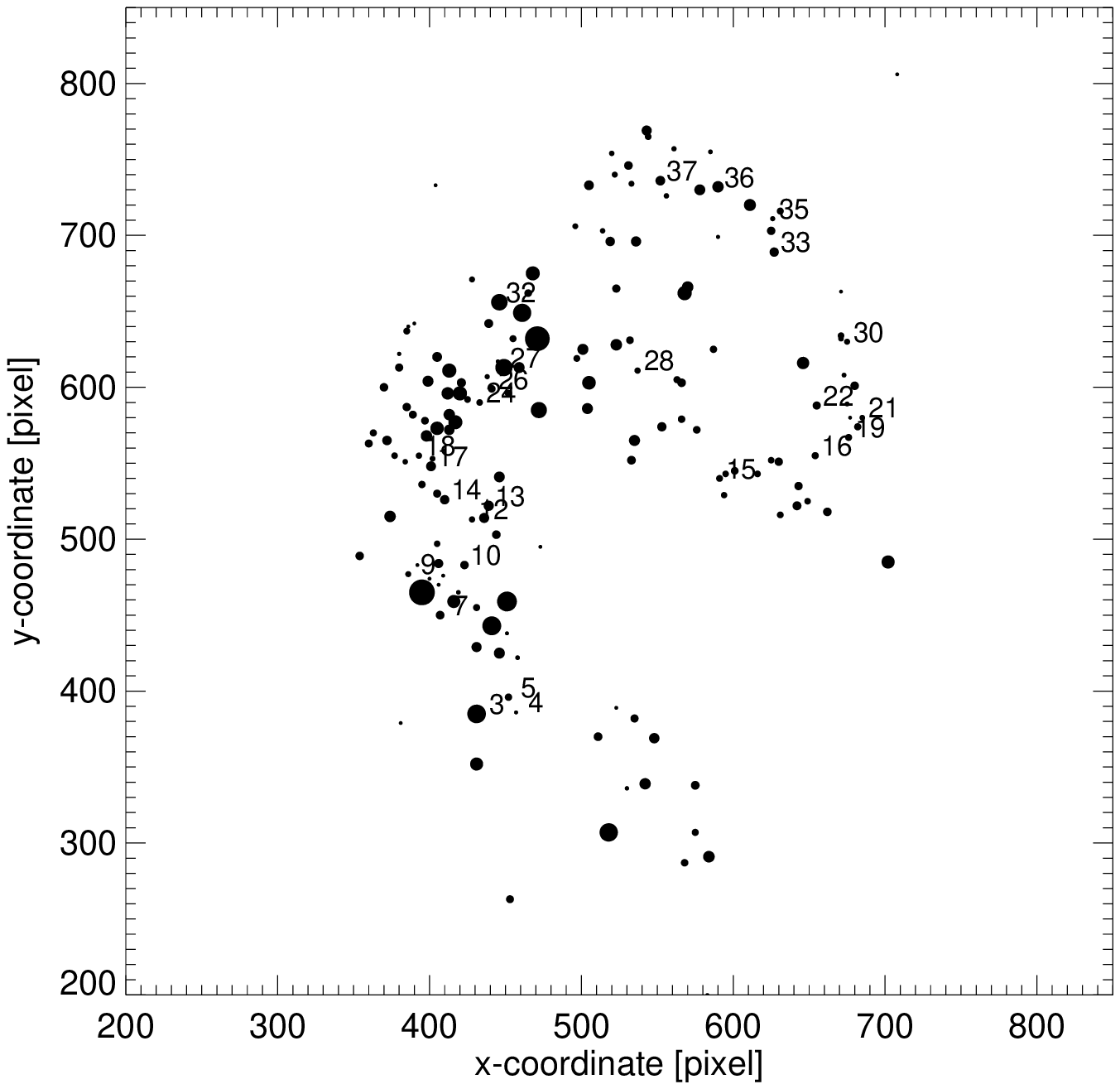}
\caption{Same as figure~\ref{clusterlumi} but with color indications
         instead of magnitudes.  Each dot refers to a near-IR source;
         the sizes of the dots scale with the ($J-K_s$) color.
         Numbers refer to the clusters at their lower left with radio
         identification in
         Table~\ref{tabcombined}. \label{clustercolor}}
\end{figure}

Table~\ref{tabreddest} lists the magnitudes of the reddest 25 clusters
with near-IR colors of $J-K_s \ge 1$.  Figure~\ref{clustercolor} best
illustrates the locations of the reddest sources.  The ($J-K_s$) color
is indicated by the size of the dots for each individual cluster.  The
overlap region shows the largest concentration of red clusters, but
red sources can also be found outside the overlap region, indicating
that star formation is not restricted to the densest region in a
global sense -- similar to what \citet{wan04} found.

\subsection{The Radio and Optical Counterparts}
VLA radio observations by \citet{nef00} have revealed numerous compact
radio sources, which can be either {\sc Hii} regions or supernova
remnants.  \citet{whi02} cross-correlated these sources with optical
{\sl HST} counterparts, which basically eliminates the sources that
are isolated supernova remnants and not associated with a young
cluster.

We have cross-correlated the combined {\sl HST} and VLA sample from
\citet{whi02} with the near-IR {\sl WIRC} sources, requiring that the
positional matches are better than $1''$ radially.  The resulting list
is shown in Table~\ref{tabcombined}.

\begin{figure}
\plotone{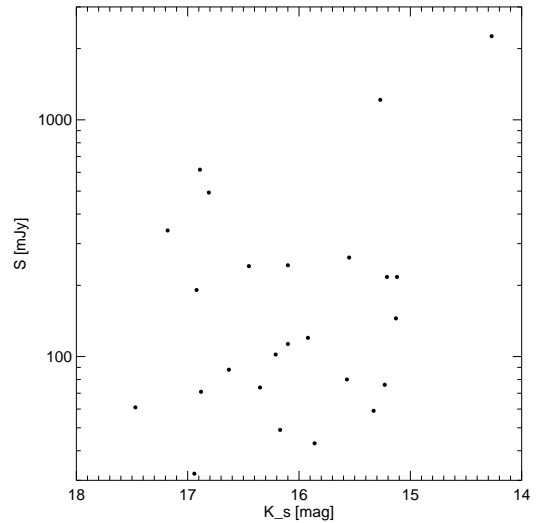}
\caption{The 6\,cm radio flux \citep{nef00} plotted against the
         infrared ($K_s$) magnitude.  The sources in this figure are
         the combined {\sl HST} and VLA sample clusters from
         \citet{whi02} with their near-IR {\sl WIRC} counterparts from
         Table~\ref{tabcombined}.  No correlation between radio and
         infrared flux is apparent. \label{radiocorr}}
\end{figure}

Figure~\ref{clustercmd} suggests that sources that have been detected with
{\sl WIRC}, {\sl HST} and VLA do not have distinct infrared colors.
In fact, they group almost symmetrically around the median ($J-K_s$)
value of the sample.  It is also apparent that the radio sources are
among the more luminous clusters of the sample.  However, except for
the two most luminous radio sources, Figure~\ref{radiocorr} shows no clear
correlation between the radio and the infrared $K_s$ brightness.

\begin{figure}[ht]
\plotone{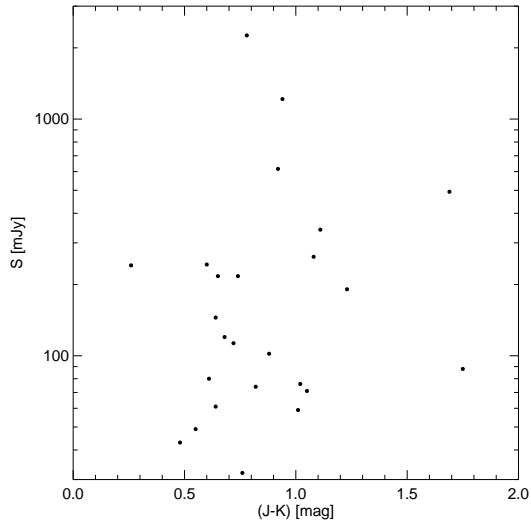}
\caption{The 6\,cm radio flux \citep{nef00} plotted against the
         infrared ($J-K_s$) color for the same sources as in
         Figure~\ref{radiocorr}. No strong correlation is
         present.\label{radiocolor}}
\end{figure}

If the color of the clusters represent a certain evolutionary
state rather than just foreground extinction one would expect some
correlation between the radio flux density and the ($J-K_s$) color.
Figure~\ref{radiocolor} shows the VLA radio flux as a function of the
infrared color.  However, no such correlation seems to be present,
suggesting that the observed optical/infared luminosity is mainly a
measure of extinction toward the young clusters, as discussed in
section~\ref{reddest}.

\begin{figure}[ht]
\plotone{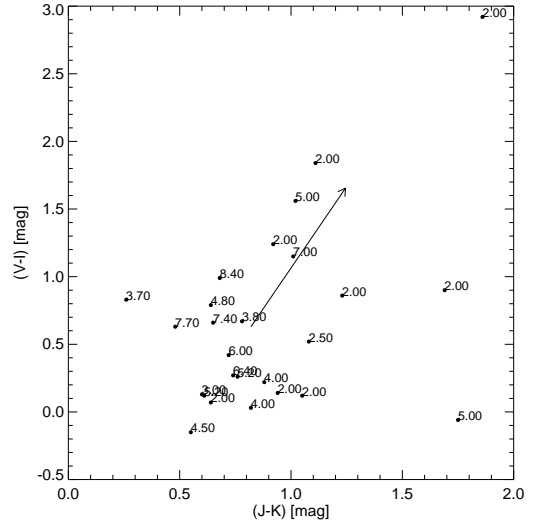}
\caption{Optical ({\sl HST}) vs. infrared ({\sl WIRC}) color-color
         diagram for the sources described in Figure~\ref{radiocorr}.
         The arrow indicates the direction of extinction; its length
         corresponds to $A_V=2.5$ and its origin is at the median
         colors of the clusters.  The numbers indicate the ages of the
         clusters in million years, according to \citet{whi02}.
         \label{optirccd}}
\end{figure}

In order to investigate if there is a correlation between optical and
infrared colors of the clusters that is {\sl not} due to extinction we
compare the the ($V-I$) vs. ($J-K_s$) colors in Figure~\ref{optirccd}.
The effect of $A_V=2.5$ foreground extinction is indicated by the
arrow.  Considering a scatter of about 0.5 magnitudes in the near-IR
due to extinction and evolution, the location of most data points in
this diagram is consistent with extinction as the dominating effect.
The most extreme outliers at the blue and red ends are statistically
not significant and may be due to mismatches between the optical
sources and the IR and/or radio sources.

Figure~\ref{optirccd} also shows for each individual cluster its age
as derived by \citet{whi02}. Within the narrow age range and the
errors associated with the age determinations, we find no obvious
correlation between age and infrared color for these young clusters.

A detailed discussion of the correlations between radio and optical
properties of the young super star clusters in the Antennae has been
given by \citet{whi02}.  However, their comparison is limited by the
rather large optical extinction toward young clusters: while the
radio-bright phase only lasts about 10~Myr it takes already about
6~Myr for a cluster to clear enough dust to reach $A_V\approx 1$
\citep{whi02}.  While they find a good correlation between the
strength of the radio flux and the $H\alpha$ emission \citet{whi02}
argue that very red clusters are not preferentially radio sources,
contrary to earlier suggestions.  Our study, providing significantly
reduced error bars on the cluster colors over a large range in
extinction, confirms the finding that the color of young clusters is
apparently not correlated with their radio brightness.  There are two
likely explanations for this finding.  First, starbursts have a
complex structure and the intrinsic color of a cluster is difficult to
disentangle from patchy foreground extinction, which can introduce a
significant scatter in Fig.~\ref{radiocolor}.  Second, the radio
emission from young clusters is predominantly produced by the most
massive stars.  Although their strong stellar winds will quickly
remove the surrounding gas and dust the associated timescales depend
strongly on the stellar density distribution and possible age spread
within the cluster \citep{bra05}.  Higher angular resolution at
near-IR wavelengths combined with accurate photometry, eventually
provided by JWST, will be of great importance.


\section{Summary} 
\label{secsummary}
 We have presented deep NIR $J$ and $K_s$ band images of the Antennae
  galaxies that cover a large area around the interaction zone.  The
 images were obtained with the new Palomar wide infrared camera {\sl
WIRC}.  We presented NIR photometry for 176 non-stellar sources within
 the observed field of view and correlated the infrared properties of
the clusters with catalogued sources from {\sl HST} and the VLA.  Our
study of 27 clusters that have been detected by {\sl WIRC}, {\sl HST}
  and the VLA shows that these young clusters cover a wide range in
     color, which is likely to be a combination of extinction and
 evolution.  The average screen extinction is about $A_V\sim 2$~mag.
 However, the reddest colors, which are mainly located in the overlap
   region, suffer from $9-10$ magnitudes in extinction.  We find no
obvious correlation between the NIR and the 6\,cm radio properties of
       these clusters.  Our study illustrates the potential of
    multi-wavelength observations and modern, sensitive wide-field
			  infrared cameras.


\acknowledgments
We would like to thank the staff at Palomar Observatory for
outstanding support.  We also thank Brad Whitmore for providing his
table of optical-radio counterparts in electronic form, and the
referee whose comments helped to significantly improve the paper.  The
{\sl WIRC} project would not have been possible without the generous
financial support from the Norris foundation, Cornell University, and
the NSF.  DMC and SSE gratefully acknowledge the support of an NSF
CAREER grant AST-0328522.



\end{document}